\documentclass[preprint,12pt]{revtex4-1}

\usepackage[brazil, english]{babel}
\usepackage{graphicx}
\usepackage{epsfig}
\usepackage{amssymb}
\usepackage{amsmath}
\usepackage{ulem}
\usepackage[T1]{fontenc} 
\usepackage[utf8]{inputenc}

\begin{document}

\title{Deconfinement phase transition in a magnetic field \\ in 2+1 dimensions from holographic models}
\author{Diego M. Rodrigues$ ^{1} $}
\email[Eletronic address: ]{diegomr@if.ufrj.br}
\author{Eduardo Folco Capossoli$^{1,2,}$}
\email[Eletronic address: ]{educapossoli@if.ufrj.br}
\author{Henrique Boschi-Filho$^{1,}$}
\email[Eletronic address: ]{boschi@if.ufrj.br}  
\affiliation{$^1$Instituto de F\'\i sica, Universidade Federal do Rio de Janeiro, 21.941-972 - Rio de Janeiro-RJ - Brazil \\
 $^2$Departamento de F\'\i sica, Col\'egio Pedro II, 20.921-903 - Rio de Janeiro-RJ - Brazil}

\begin{abstract} 
Using two different models from holographic quantum chromodynamics (QCD) we study the deconfinement phase transition in $2+1$ dimensions in the presence of a magnetic field. Working in 2+1 dimensions lead us to {\sl exact} solutions on the magnetic field, in contrast with the case of 3+1 dimensions where the solutions on the magnetic field are perturbative. As our main result we predict a critical magnetic field $B_c$ where the deconfinement critical temperature vanishes. For weak fields meaning $B<B_c$ we find that the critical temperature decreases with increasing magnetic field indicating an inverse magnetic catalysis (IMC). On the other hand, for strong magnetic fields $B>B_c$ we find that the critical temperature raises with growing field showing a magnetic catalysis (MC).  These results for IMC and MC are in agreement with the literature. 
\end{abstract}


\maketitle


\newcommand{\limit}[3]
{\ensuremath{\lim_{#1 \rightarrow #2} #3}}

\section{Introduction}   

The deconfinement phase transition in quantum chromodynamics (QCD) still remains an open and intriguing problem, since the standard perturbative method  does not work due to the strong coupling regime at low energies. The usual  approach to deal with this non-perturbative issue is lattice QCD where one finds a critical temperature $T_c$ which characterises the  deconfinement phase transition. In particular, the presence of a magnetic field $B$ modifies this scenario.  It has been shown recently \cite{Bali:2011qj} that weak magnetic fields imply a decreasing $T_c$, an effect known as inverse magnetic catalysis (IMC). Furthermore, it is expected that for strong magnetic fields $T_c$ increases with $B$, meaning a magnetic catalysis (MC). The MC/IMC studies are usually concerned with chiral symmetry breaking and/or deconfinement phase transition. 

Currently, many works have dealt with MC/IMC using a holographic approach based on the AdS/CFT correspondence. This correspondence or duality makes it possible to relate strong coupling theory in flat Minkowski space with weak coupling supergravity in anti-de Sitter space (AdS) in a higher dimensional space \cite{Maldacena:1997re, Gubser:1998bc, Witten:1998qj}. Among these works we can mention \cite{Evans:2010xs, Alam:2012fw,  Filev:2010pm, Callebaut:2011zz, Bolognesi:2011un} where they study the MC problem, while \cite{Preis:2010cq, McInnes:2015kec, Evans:2016jzo, Mamo:2015dea, Li:2016gfn, Dudal:2015wfn} discuss IMC effects in different holographic models. Note that in all of these 3+1 dimensional models the gravitational solutions on the magnetic field are perturbative. 

Here in this work we study deconfinement phase transition  in 2+1 dimensions in the presence of an external magnetic field $B$ within two different holographic AdS/QCD models. We find the IMC and MC pictures for the deconfinement phase transition and obtain an intriguing critical magnetic field $B_c$ for which the  critical temperature $T_c$ vanishes. The advantage of working in 2+1 dimensions is that the system of equations are simpler than the 3+1 dimensional case, leading us to some {\sl exact} solutions where we can obtain the IMC/MC transition at a critical value of $B$=$B_c$. In ref. \cite{Bolognesi:2011un} the case of 2+1 dimensions was studied for the case of the MC on the fermion condensate. 

The holographic  models that we use are known as the hard \cite{Polchinski:2001tt, BoschiFilho:2002vd}   and soft wall  \cite{Karch:2006pv, Colangelo:2007pt}.  Such models were successful in predicting the deconfinement phase transition and its critical temperature $T_c$ in the absence of a magnetic field \cite{Herzog:2006ra, BallonBayona:2007vp}. 
These holographic models appeared after the proposal of the AdS/CFT correspondence, which provides an approach to deal problems  out of the  perturbative regime of QCD or other strongly interacting systems.

This work is organized as follows: in section \ref{Geo} we review the Einstein-Maxwell Theory in 4 dimensions and the geometric set up in the presence of an external magnetic field. In section \ref{Models} we describe the holographic models used and compute the corresponding on-shell actions for both thermal and black hole AdS spaces. Then, in section \ref{DPT}, we present our results for the deconfinement phase transition in the hard and soft wall models in the presence of an external magnetic field and obtain the critical magnetic field $B_c$. Finally, in section \ref{discussion} we present our last comments and conclusions. 

\section{Einstein-Maxwell Theory in 4 dimensions}\label{Geo}

Here, we start with holographic models defined in $AdS_4$ such that the dual field theory in Euclidean space lives in 3 dimensions. 
 The full gravitational background is the eleven-dimensional supergravity on $ AdS_4\times S^7 $. The dual field theory is the low-energy theory living on $ N$ $M2 $-branes on $ \mathbb{R}^{1,2} $, with $ \mathcal{N}=8 $ $ SU(N) $ Super-Yang-Mills theory in the large $ N $ limit \cite{Aharony:1999ti}. 
Via Kaluza-Klein dimensional reduction, the supergravity theory on $ AdS_4\times S^7 $ may be consistently truncated to Einstein-Maxwell Theory on $ AdS_4$ \cite{Herzog:2007ij}. The action for this theory, in Euclidean signature
%
is given by 
\begin{eqnarray}\label{AdS4Action}
S_{\mathrm{Ren}} = -\dfrac{1}{2\kappa^2_4}\int d^{4}x \sqrt{g}\left(R -2\Lambda - L^{2}F_{MN}F^{MN}\right) - \dfrac{1}{\kappa^2_4}\int d^{3}x \sqrt{\gamma}\left(K + \dfrac{4}{L}\right).
\end{eqnarray}
where $ \kappa^2_4 $ is the 4-dimensional coupling constant, which is proportional to the 4-dimensional Newton's constant $( \kappa^2_4\equiv8\pi G_4 )$, ${R} $ is the Ricci scalar and $ \Lambda $ is the negative cosmological constant which, for $ AdS_4$, are given by $R = -{12}/{L^2}$, and $\Lambda = -{3}/{L^2}$, 
respectively. $ L $ is the radius of $ AdS_4$ and $ F_{MN} $ is the Maxwell field.  
The second integral corresponds to the surface and counter-terms in which  $ \gamma $ is the determinant of the induced metric $\gamma_{\mu\nu}$ on the boundary, and $ K = \gamma^{\mu\nu}K_{\mu\nu} $ is the trace of the extrinsic curvature $K_{\mu\nu} $ which gives the Gibbons-Hawking surface term \cite{Gibbons:1976ue}. The last term is a counter-term needed to cancel the UV divergences ($ z\rightarrow0 $) of the bulk action.

The field equations coming from the bulk action \eqref{AdS4Action} are \cite{Herzog:2007ij}
\begin{eqnarray}
R_{MN} &=& 2L^2\left( F_{M}^{P}F_{NP}-\dfrac{1}{4}g_{MN}F^2\right)  - \dfrac{3}{L^2}g_{MN}, \label{FieldEquations}
\end{eqnarray}
together with the Bianchi identities $\nabla_{M}F^{MN} = 0$. 
The ansatz for the metric to solve these equations is given by
\begin{eqnarray}
ds^2 &=& \dfrac{L^2}{z^2}\left( f(z)d\tau^2 + \dfrac{dz^2}{f(z)} + dx^2_1 + dx^2_2\right), \label{AnsatzMetrica}
\end{eqnarray}
 in Euclidean signature with a compact time direction, $0\leq\tau\leq\beta$, with $ \beta = \frac{1}{T} $, and 
 $f(z)$ is a function to be determined in the following. 
The background magnetic field is chosen such that $F = B\, dx_{1}\wedge dx_{2}$, which implies $ F^2 = {2B^{2}z^4}/{L^4}$. Note that the magnetic field remains finite at the $ AdS_4 $ boundary ($ z\rightarrow0 $). To see this let's consider the vector potential, which is a 1-form $ A $ such that $ F = dA $. So,
$ A = \dfrac{B}{2}(x_{1}dx_{2}-x_{2}dx_{1}) $. 
  Thus, we can treat it as an external background magnetic field \cite{Hartnoll:2007ai}.

Using the ansatz \eqref{AnsatzMetrica}  the field equations \eqref{FieldEquations} are simplified and given by
\begin{eqnarray} 
z^2f''(z)-4zf'(z)+6f(z)-2B^{2}z^4-6 = 0, \label{FieldequationSimplified1} \cr 
zf'(z)-3f(z)-B^{2}z^4+3 = 0. 
\end{eqnarray} 
\noindent 
The two {\sl exact} solutions of \eqref{FieldequationSimplified1}  that we found are given by
\begin{eqnarray}
f_{Th}(z) &=& 1 + B^{2}z^4 \label{fT} \cr 
f_{BH}(z) &=& 1 + B^{2}z^3(z-z_H) - \dfrac{z^3}{z^{3}_H} 
\end{eqnarray} 
The first solution, $ f_{Th}(z) $, corresponds to the thermal $ AdS_4 $ with an external background magnetic field. The second solution, $ f_{BH}(z) $, corresponds to a black hole in $ AdS_4 $ also in the presence of a background magnetic field, and where $ z_H $ is the horizon position, such that $ f_{BH}(z=z_H)=0 $. One can note the these two solutions indeed satisfy both differential equations \eqref{FieldequationSimplified1}. This is in contrast with the 3+1 dimensional case where only perturbative solutions on the magnetic field $B$ are found.

\section{On-shell Euclidean Actions} \label{Models}

\subsection{Hard wall}

The hardwall model \cite{Polchinski:2001tt,BoschiFilho:2002vd} consists in introducing a hard cut-off in the background geometry in order to break conformal invariance. 
The introduction of a cut-off $ z_{max} $ in this model implies that 
$ 0\leqslant z\leqslant z_{max}$, 
where $ z_{max} $ can be related to the mass scale of the boundary theory. For instance, in $ 4 $ dimensions $ z_{max} $ is usually related with energy scale of QCD \cite{BoschiFilho:2005yh,Rodrigues:2016cdb} by 
$ z_{max} \sim \dfrac{1}{\Lambda_{QCD}}$. 
Moreover we have to impose boundary conditions in $ z=z_{{max}} $. 


In the hard wall model, the free energy for the thermal AdS$_4$, from the action  \eqref{AdS4Action}, is given by (see \cite{DCB2017} for details): 
\begin{equation}
S_{Th} = \dfrac{\beta'\mathcal{V}_{2}L^2}{\kappa^2_4}\left(-\frac{1}{z_{max}^3} + B^2z_{max} + \mathcal{O}(\epsilon)\right)\,,
\end{equation} 
where $ \mathcal{V}_2\equiv \iint dx_{1}dx_{2} $, $\beta'$ is the corresponding period, and $\epsilon$ is an UV
regulator.


On the other hand, for the black hole case one gets
 the free energy 
\begin{equation}
S_{BH} = \dfrac{\beta\mathcal{V}_{2}L^2}{\kappa^2_4}\left( -\dfrac{1}{2z_{H}^3} + \dfrac{3 B^2 z_{H}}{2} +\mathcal{O}(\epsilon)\right) \,,
\end{equation}
where $\beta$ is associated with the Hawking temperature. 

Now we have to compute the free energy difference, $\Delta S$, defined by
$ \Delta S = \limit{\epsilon}{0}(S_{BH}-S_{Th})$. 
Since we are comparing the two geometries at the same position $ z=\epsilon\rightarrow0 $ we can choose $ \beta' $ such that $ \beta' = \beta\sqrt{f(\epsilon)} = \beta $ \cite{Herzog:2006ra, BallonBayona:2007vp}, since $ f(\epsilon) = 1 + \mathcal{O}(\epsilon^3) $ when $ \epsilon\rightarrow0 $, with $ f(z) $ given by the second equation in \eqref{fT}. Therefore, with this choice, we have that the free energy difference for the hardwall model is given by
\begin{equation}\label{FenergyDifHard}
\Delta S_{HW} = \dfrac{\beta\mathcal{V}_{2}L^2}{\kappa^2_4}\left(\dfrac{1}{z_{max}^3} -\dfrac{1}{2z_{H}^3} + B^2 \left(\dfrac{3z_{H}}{2}-z_{max}\right)\right).
\end{equation} 
For $ B = 0 $, this equation corresponds to the 3-dimensional version of \cite{Herzog:2006ra}.  

\subsection{Soft wall}

For the soft wall model \cite{Colangelo:2007pt,Karch:2006pv} we consider the following 4-dimensional action
\begin{equation} \label{SoftwallAction}
S_{SW} = -\dfrac{1}{2\kappa^2_4}\int d^{4}x \sqrt{g}\,e^{-\Phi(z)}\left(\mathcal{R} -2\Lambda - L^{2}F_{\mu\nu}F^{\mu\nu}\right)  - \dfrac{1}{\kappa^2_4}\int d^{3}x \sqrt{\gamma}\left(K + \dfrac{4}{L} - \dfrac{3\Phi}{L}\right), 
\end{equation}
where $ \Phi(z) = kz^2 $ is the dilaton  field, which has non-trivial expectation value. In this work we are assuming that the dilaton field does not backreact on the background geometry. Moreover, as in \cite{Herzog:2006ra}, we assume that our metric ansatz \eqref{AnsatzMetrica} satisfies the equations of motion for the full theory with $ f(z) $ given by \eqref{fT} for both thermal and black hole in $ AdS_4 $. 
One can note that we included one more term on the boundary action compared to \eqref{AdS4Action}, due to the dilaton field in this soft wall model. 

 The free energy for the thermal $ AdS_4 $, in the soft wall model is given by
\begin{equation}
S_{Th} = \dfrac{\beta'\mathcal{V}_{2}L^2}{\kappa^2_4}\left(\dfrac{\sqrt{\pi }(B^2 + 4k^2)}{2 \sqrt{k}} + \mathcal{O}(\epsilon)\right).
\end{equation}


On the other hand, the free energy for the $ AdS_4 $ black hole for the soft wall model, is 
\begin{eqnarray}
S_{BH} = \dfrac{\beta\mathcal{V}_{2}L^2}{\kappa^2_4} \bigg(&\dfrac{1}{2z_{H}^3}& + \dfrac{e^{-k z_{H}^2} \left(2kz_{H}^2-1\right)}{z_{H}^3} + \dfrac{B^2z_{H}}{2} + \nonumber \\
&+&\dfrac{\sqrt{\pi }\left(B^2+4 k^2\right) \text{erf}\left(\sqrt{k}z_{H}\right)}{2\sqrt{k}} + \mathcal{O}(\epsilon)\bigg)\,,
\end{eqnarray}
where $\text{erf}(z)$ is the error function. 

Therefore, taking into account the same argument which led to $ \beta'=\beta\sqrt{f(\epsilon)} =\beta $ in the hardwall model, the free energy difference, $\Delta S$, for the softwall model is given by
\begin{equation}\label{FenergyDifSoft}
\Delta S_{SW} =  \dfrac{\beta\mathcal{V}_{2}L^2}{\kappa^2_4}\left(\dfrac{1}{2z_{H}^3} + \dfrac{e^{-kz_{H}^2} \left(2kz_{H}^2-1\right)}{z_{H}^3} + \dfrac{B^2z_{H}}{2} - \dfrac{\sqrt{\pi }\left(B^2+4 k^2\right)\text{erfc}\left(\sqrt{k}z_{H}\right)}{2\sqrt{k}}\right), 
\end{equation}
where erfc$ (z) $ is the complementary error function, defined as $\text{erfc}(z) = 1 - \text{erf}(z) $.

\section{Deconfinement Phase Transition}\label{DPT}

Following Hawking and Page \cite{Hawking:1982dh} and Witten \cite{Witten:1998zw}, we study the deconfinement phase transition imposing 
$ \Delta S(z_{Hc},B) = 0, $ 
where $ z_{Hc} $ is the critical horizon, from which we calculate the critical temperature through the formula $ T_c = {|f'(z=z_{Hc})|}/{4\pi},$ 
where $ f(z) $ is the horizon function given by \eqref{fT}.

In the hard wall model with $B=0$ and  from \eqref{FenergyDifHard} we find that the deconfinement phase transition occurs at $2 z_{Hc}^3=z_{max}^3$ resulting in 
the critical temperature $ T_c(B=0) \approx {0.3} / {z_{max}} $, 
which is the analogue  in $ (2+1) $ dimensions of \cite{Herzog:2006ra,BallonBayona:2007vp}. 
In order to fix the cut off $ z_{max} $ we use Neumann boundary condition which gives $ J_{1/2}(m \, z_{max})=0 $ so that 
$ z_{max}= 3.141/m $, where $ m $ is the lightest scalar glueball mass 
$ m_{0^{++}}/ \sqrt{\sigma}$ =  4.37 for SU(3) in (2+1) dimensions where $\sqrt{\sigma}$ is the string tension \cite{Teper:1998te, Athenodorou:2016ebg}. 
Then, one can compute $ z_{max} $ and the critical temperature, $ T_c $, in units of the string tension for $B=0$: 
\begin{equation}\label{TcResultB=0Hard}
\dfrac{T_c(B=0)}{\sqrt{\sigma}}  =  0.42 \qquad  {\rm (hard\, wall)}\,.
\end{equation}

For the soft wall model, for $ B = 0 $,  there is a phase transition when $ \sqrt{k}\, z_{Hc} = 0.60 $ which gives the critical temperature 
$ T_{c}(B=0) = 0.40 \sqrt{k},$ 
consistent with the treatment presented in \cite{Herzog:2006ra,BallonBayona:2007vp} for $ B=0 $ in one higher dimension. 
In order to fix the value of $ k $ we consider the soft wall model in 4 dimensions  so that we have 
$ m_{n}^2 = \left(4n + 6\right)k$ (see \cite{DCB2017} for details). 
Using the mass for the lightest glueball in $ (2+1) $ dimensions from the lattice \cite{Athenodorou:2016ebg} and setting $ n=0 $, we can fix the dilaton constant 
$ k =$ 3.18 for the SU(3), in units of the string tension squared.
Therefore, the critical temperature, $ T_c(B=0) $, in units of the string tension,  is given by
\begin{equation}\label{TcResultB=0Soft}
\dfrac{T_c(B=0)}{\sqrt{\sigma}}  =  0.71\qquad  {\rm (soft \, wall)}\,.
\end{equation}

\begin{figure}
	\centering
	\includegraphics[scale=0.45]{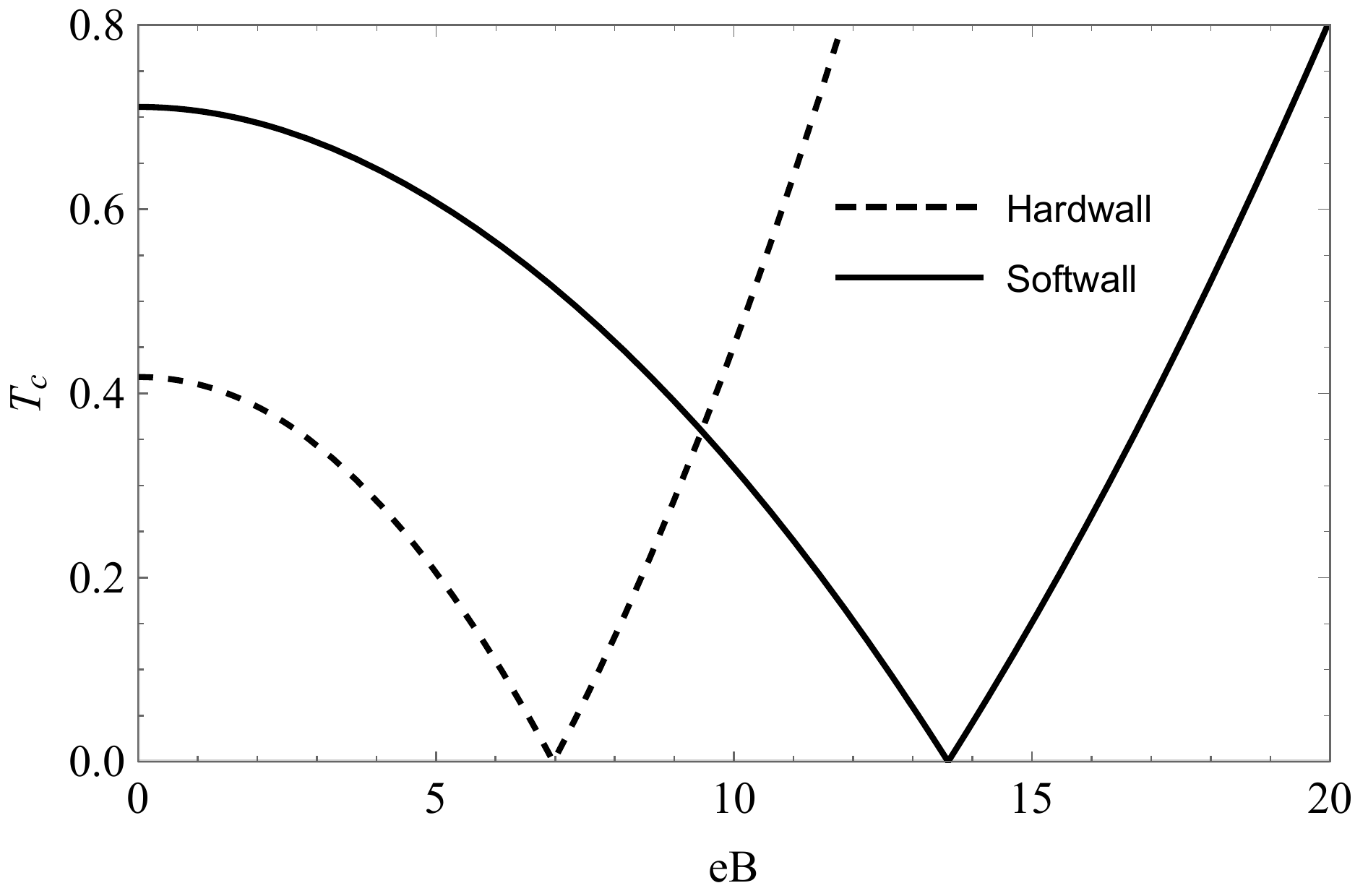}
	\caption{Critical temperature (in units of string tension) as a function of the magnetic field $eB$ (in units of string tension squared).}
	\label{HardSoft}
\end{figure}

On the other hand, for $ B\neq0 $ from \eqref{FenergyDifHard} (hard wall), and \eqref{FenergyDifSoft} (soft wall), 
the numerical results for the critical temperature as a function of the magnetic field, $T_{c}(B)$, is shown in Figure \ref{HardSoft}, for both models. One can see from this figure that we have a phase in which the critical temperature, $T_{c}(B)$, decreases with increasing magnetic field $B$, indicating an {\sl inverse magnetic catalysis} (IMC). Furthermore, we also predict a phase in which the critical temperature, $T_{c}(B)$, increases with increasing magnetic field $B$, indicating a {\sl magnetic catalysis} (MC). 

The magnetic and inverse magnetic catalysis we have found for these models are separated by a critical magnetic field, $ B_c $. The values of the critical magnetic fields found in these models, in units of the string tension squared, are the following
\begin{equation} \label{CriticalMagneticFieldHard}
\dfrac{eB_c}{\sigma} =  6.97 \qquad  {\rm (hard\, wall)}\,;
\end{equation}
\begin{equation} \label{CriticalMagneticFieldSoft}
\dfrac{eB_c}{\sigma} =  13.6 \qquad  {\rm (soft \, wall)}\,.
\end{equation}

\begin{figure}
	\centering
	\includegraphics[scale=0.45]{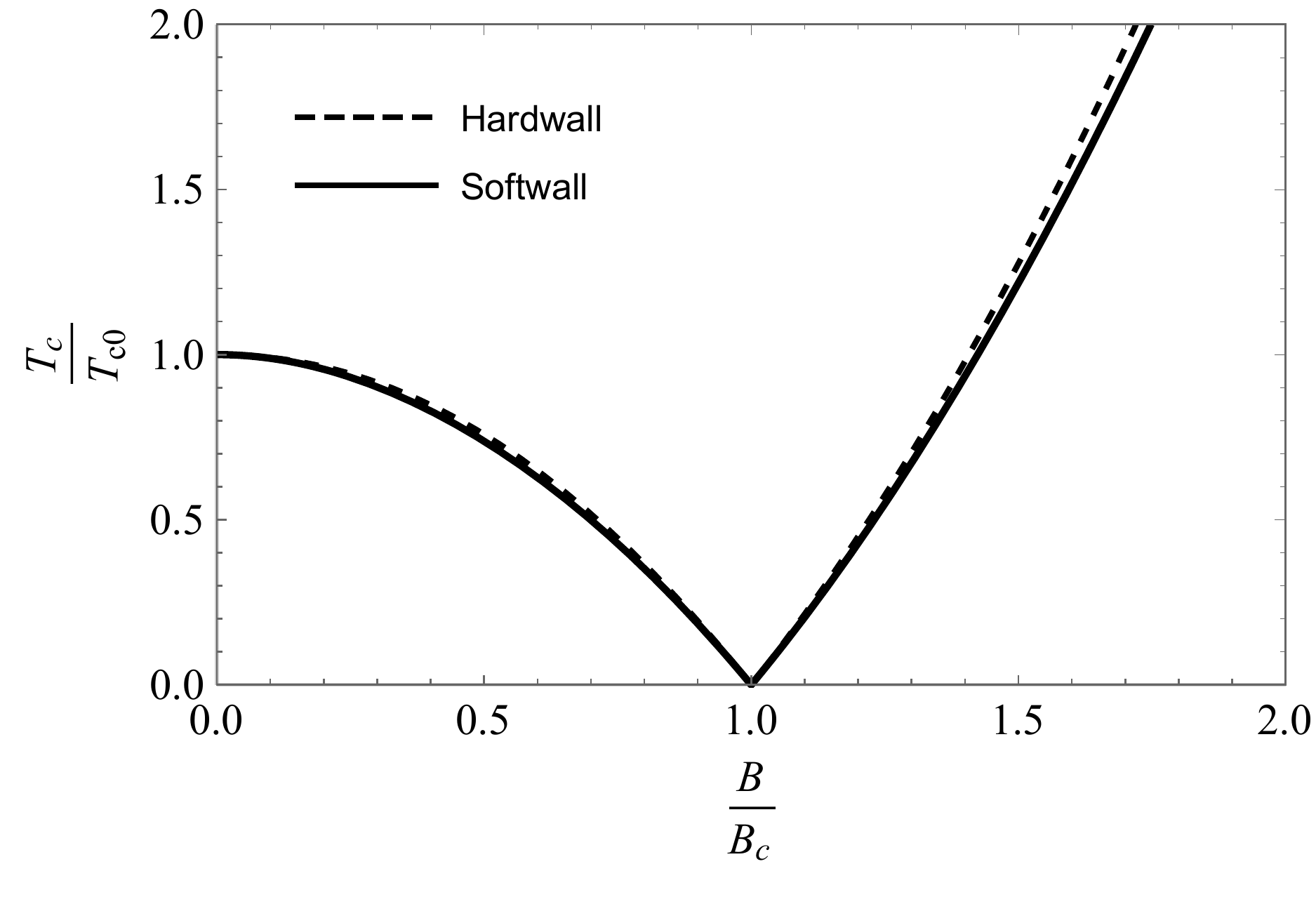}
	\caption{The ratio of the critical temperatures as a function of the ratio of the magnetic fields.}
	\label{hardsoftAdim}
\end{figure}

In Figure \ref{hardsoftAdim} we show the plot of the normalized critical temperature, $ T_c/T_{c_0} $, as a function of $ B/B_c $ for both models, where $ T_{c_0}\equiv T_c(B=0) $ and $ B_c $ is the critical magnetic field \eqref{CriticalMagneticFieldHard}, and  \eqref{CriticalMagneticFieldSoft}. 

\section{Discussions}\label{discussion}

The IMC  has been observed in lattice QCD \cite{Bali:2011qj} for $ eB\lesssim $1\ GeV$^2 $. Since then many holographic approaches reproduced this behavior in both deconfinement and chiral phase transition contexts within this range of magnetic field, see for instance \cite{Mamo:2015dea,Dudal:2015wfn,Evans:2016jzo,Li:2016gfn}. 

However, in many of these approaches the problem could only be solved perturbatively in $B$, while in our results  in (2+1) dimensions there is no restriction for the values or range of the magnetic field. This is in contrast with the 3+1 dimensional case where only perturbative solutions on the magnetic field $B$ are found. 
Since we are working in $ (2+1) $ dimensions, physical quantities such as the critical temperature, $T_c$, magnetic field, $B$, and critical magnetic field, $B_c$, are not measured in GeV or MeV.  
Instead we used the string tension  $ \sqrt{\sigma} $ as the basic unit for our physical quantities, as is the case in lattice simulations \cite{Athenodorou:2016ebg,Teper:1998te,Meyer:2003wx}. 

In conclusion, we emphasize that the critical magnetic field found here is an unexpected result since in 3+1 dimensional QCD there is evidence that the deconfinement (and chiral) transition is a cross over \cite{Bali:2011qj}.

\vspace{12pt}
\noindent {\bf Acknowledgments:} 
We would like to thank Luiz F. Ferreira, Adriana Lizeth Vela, Renato Critelli, R\^omulo Rogeumont, and Marco Moriconi for helpful discussions during the course of this work. We also thank Elvis do Amaral for the help with numerical solutions. We would also like to thank Michael Teper for useful correspondence. D.M.R is supported by Conselho Nacional de Desenvolvimento Cient\'\i fico e Tecnol\'ogico (CNPq), E.F.C. is partially supported by PROPGPEC-Col\'egio Pedro II, and H.B.-F. is partially supported by CNPq.

 \end{document}